\documentstyle[12pt,aaspp4, epsf]{article}

\def\Mdot{\hbox{$\dot {M}$}}

\def\Lsun{\hbox{\it L$_\odot$}}

\def\Msun{\hbox{\it M$_\odot$}}

\def\Msunyr{\hbox{\it M$_\odot\,$yr$^{-1}$}}

\def\kms{\hbox{km$\,$s$^{-1}$}}

\def\AK{\hbox{\it A$_{\rm K}$}}

\def\simgr{\mathrel{\hbox{\rlap{\hbox{\lower4pt\hbox{$\sim$}}}\hbox{$>$}}}}
\def\HH{H{\sc ii}}	

\makeatletter
\def\jnl@aj{AJ}
\ifx\revtex@jnl\jnl@aj\fi
\makeatother

\received{19 February 1999}
\revised{REVISION DATE}
\accepted{24 June 1999}
\journalid{VOL}{JOURNAL DATE}
\articleid{START PAGE}{END PAGE}
\paperid{MANUSCRIPT ID}
\cpright{TYPE}{YEAR}
\ccc{CODE}

\lefthead{Figer et al.}
\righthead{Pistol}
\begin{document}

\title{High Resolution Infrared Imaging and Spectroscopy \\ of the Pistol Nebula: Evidence for 
Ejection\footnote{Based on observations with the 
NASA/ESA Hubble Space Telescope, obtained
at the Space Telescope Science Institute, which is operated by the Association of Universities
for Research in Astronomy, Inc. under NASA contract No. NAS5-26555.}}

\author{Donald F. Figer\altaffilmark{2,3}, Mark Morris\altaffilmark{2,8},  \\
T. R. Geballe\altaffilmark{4}, 
R. Michael Rich\altaffilmark{2}, Eugene Serabyn\altaffilmark{5}, \\ 
Ian S. McLean\altaffilmark{2}, R. C. Puetter\altaffilmark{6}, Amos Yahil\altaffilmark{7}}

\authoremail{figer@astro.ucla.edu}

\altaffiltext{2}{Division of Astronomy, 
Department of Physics \& Astronomy, University of California, Los Angeles, 
405 Hilgard Avenue, Los Angeles, CA 90095-1562; figer@astro.ucla.edu, morris@astro.ucla.edu, rmr@astro.ucla.edu,
mclean@astro.ucla.edu}
\altaffiltext{3}{Space Telescope Science Institute, 3700 San Martin Drive, Baltimore, MD 21218}
\altaffiltext{4}{Gemini Observatory, 670 N. A'ohoku Place, Hilo, HI  96720; tgeballe@gemini.edu}
\altaffiltext{5}{JPL 171-113, 4800 Oak Grove Dr., Pasadena, CA 91109; eserabyn@huey.jpl.nasa.gov}
\altaffiltext{6}{Center for Astrophysics and Space Sciences, University of California, San Diego, 9500 Gilman
Drive, La Jolla, CA 92093-0111; rpuetter@ucsd.edu}
\altaffiltext{7}{Department of Physics and Astronomy, State University of New York at Stony Brook, Stony Brook, NY 11794-3800; Amos.Yahil@sunysb.edu}
\altaffiltext{8}{Institut d'Astrophysique de Paris, 98 bis Blvd Arago, 75014 Paris, France}

\begin{abstract}
We present new infrared images, obtained with the {\it Hubble Space Telescope} ({\it HST}) 
Near-infrared Camera and Multi-object Spectrometer (NICMOS), and Br-$\alpha$ (4.05~\micron) spectroscopy, 
obtained using CGS4 on UKIRT, of the Pistol Star and its associated nebula.  
We find strong evidence to support the hypothesis that the Pistol Nebula was ejected 
from the Pistol Star. 
The Pa-$\alpha$ (1.87~\micron) NICMOS image shows that the nebula completely surrounds the Pistol Star, although the 
line intensity is much stronger on its northern and western edges. 
The Br-$\alpha$ CGS4 spectra show the classical
ring-like signature of quasi-spherical expansion. The blueshifted emission (V$_{\rm max}$ $\approx$ $-$60~\kms) is
much weaker than the redshifted emission (V$_{\rm max}$ $\approx$ $+$10~\kms), where the velocities are with
respect to the velocity of the Pistol Star; further, the redshifted emission spans a very narrow
range of velocities, i.e., it appears ``flattened'' in the position-velocity diagram. 
These data suggest that the nebula was ejected from the star several thousand 
years ago, with a velocity between the current terminal velocity of the stellar wind (95~\kms) and 
the present expansion velocity of gas in the outer shell of the nebula (60~\kms). 
The Pa-$\alpha$ image reveals several emission-line stars in the region, including 
two newly-identified emission-line stars north of the Pistol Star, both of which are likely
to be the hottest known stars in the Galactic center with spectral types earlier than WC8 
and T$_{\rm eff}$~$>$~50,000 K). The presence of these stars, the morphology of the Pa-$\alpha$ emission, 
and the velocity field in the gas suggest that the
side of the nebula furthest from us is approaching, and being ionized by, the hot stars of the Quintuplet, and
that the highest velocity redshifted gas has been decelerated by 
winds from the Quintuplet stars. We also discuss the possibility that the nebular gas might be
magnetically confined by the ambient magnetic field delineated by the nearby nonthermal filaments.
\end{abstract}

\keywords{stars: evolution --- circumstellar matter --- stars: mass-loss --- 
stars: high mass --- Galaxy: center --- ISM: individual (G0.15-0.05)}

\section{Introduction}

The Pistol \ion{H}{2} region\markcite{yzm} (Yusef-Zadeh \& Morris 1987) surrounds one of the most
luminous stars known, the ``Pistol Star''\markcite{f98} (Figer et al.\ 1998). 
Located near the Galactic center ({\it l,b}~=~$+0.15, -0.05)$,
the nebula appears to be elongated parallel to, and
precisely bounded by, two prominent filaments of the Radio Arc in the Galactic center.
The apparent coincidence of the Pistol Nebula and the nearby Sickle \ion{H}{2} region (G0.18$-$0.04)
with the nonthermal radio filaments has prompted several groups to consider
that gas in the \ion{H}{2} regions might be 
ionized by the relativistic particles in the filaments\markcite{yzm,sg91} 
(Yusef-Zadeh \& Morris 1987; Serabyn \& G\"usten 1991). Moneti, Glass, \& Moorwood\markcite{mgm94} (1994)
first identified the ``Pistol Star,'' their ``serendipitous source,'' 
noting its strong Br-$\gamma$ emission. Cotera et al.\markcite{cot94} (1994) 
independently identified the same star as ``Pistol Source A,'' suggesting that it might be
responsible for ionizing the Pistol Nebula\markcite{har94} (see also, Harris et. al.\ 1994).
Figer, McLean, \& Morris\markcite{fmm95} (1995; hereafter FMM95) and 
Figer, McLean, \& Morris\markcite{f99} (1999; hereafter FMM99) proposed that the star is
an LBV and that 
the nebula is photoionized by the nearby large population of extremely
massive, luminous stars, some of which are Wolf-Rayet stars, in agreement with
Timmermann et al.\markcite{tim96} (1996). FMM95 further proposed that the Pistol Nebula is 
circumstellar ejecta from the Pistol Star; note that the Pistol Star has the same
line-of-sight velocity as the Pistol Nebula\markcite{f95;ymv89} (Figer 1995; Yusef-Zadeh, 
Morris, \& van Gorkom 1989). 

F98 find that the Pistol Star is extremely luminous\markcite{fmm95,cote96} 
(L~$>$~10$^{6.6}$~\Lsun; also see FMM95 and Cotera et al.\ 1996), with no
evidence of a comparably bright companion to within 14 mas (110 AU). 
F98 argue that the star is a Luminous Blue Variable\markcite{cont84,hd94} 
(``LBV''; Conti 1984; Humphreys \& Davidson 1994), given the star's 
position in the HR diagram, placement within the Pistol Nebula,
and near-infrared spectrum\markcite{fmm95} (also see FMM95). Most stars in the LBV class are
surrounded by circumstellar ejecta\markcite{nota95} (Nota et al.\ 1995), and
Conti\markcite{conti} (1997) has suggested that 
these nebulae should be regarded as identifying characteristics of the class.
Recent photometric monitoring by Glass et al.\markcite{glass99} (1999) has confirmed that the 
Pistol Star is photometrically variable by $\approx$0.5 mag. in the {\it K}-band.

This paper provides new evidence that the Pistol Nebula is photoionized circumstellar
ejecta emitted by the Pistol Star, although we argue that the nebula is primarily ionized
by nearby hot stars elsewhere in the Quintuplet cluster. We present {\it HST}/NICMOS 
Pa-$\alpha$ (\ion{H}{1} 3-4, 1.87~\micron) images which reveal the Pistol
nebula in new detail and show that the nebula completely surrounds the 
Pistol Star.  We present Br-$\alpha$ (\ion{H}{1} 4-5, 4.05~\micron) spectroscopy which shows that the ionized gas 
in the nebula has a velocity structure characteristic of quasi-spherical 
expanding systems, such as planetary nebulae.  We propose
that the Pistol Star is an extremely luminous LBV surrounded
by a massive ejection nebula which is primarily ionized externally by nearby hot stars and which is
physically interacting with the strong winds of these stars.

\section{Observations}

\subsection {NICMOS Imaging}

The imaging data were obtained using {\it HST}/NICMOS. 
A 2$\times$2 NIC2 mosaic was obtained on UT 1997 September 13/14 in the F187N ($\lambda_{\rm center}$
=~1.87~\micron, Pa-$\alpha$) 
and F190N ($\lambda_{\rm center}$
=~1.90~\micron, continuum) filters with 19\farcs05 spacing between images;
the plate scale was 0\farcs076 pixel$^{-1}$ (x) by 0\farcs075 pixel$^{-1}$ (y), in detector coordinates.
The STEP256 sequence was used in the MULTIACCUM read mode with 11 reads, i.e., the exposure time
was $\approx$256 seconds per image. The pattern was centered on RA 17$^{\rm h}$ 43$^{\rm m}$ 5$\fs$0,  
DEC $-$28$^{\arcdeg}$ 48$^{\arcmin}$ 54$\farcs$87 (B1950) with an orientation of 
$-$134$\fdg$7. The names of the data files are given in Table 1.
The data were reduced with STScI pipeline routines, calnica and calnicb, using the calibration files listed in Table 2. 

The color composite image in Figure 1 was produced with calibration files available at the time
of the observations. In this figure, the difference of the
F187N and F190N images was coded as red, the F190N image as blue, and the F187N image as green. 
A re-calibration was used to produce the other figures and to extract
the photometry, as described below. 
The second calibration used more recent files, most notably, on-orbit flat field
images. Figure 2 shows a negative greyscale image of the difference frame (F187N$-$F190N).

\subsection{Spectroscopy}
 
Spectra in the Br-$\alpha$ region were obtained with the cooled
grating spectrometer, CGS4, on the United Kingdom Infrared Telescope (UKIRT) on
UT 1998 July 5, July 10, and August 3. The echelle grating and the 300 mm focal length
camera were used to produce a plate scale of 0\farcs45 $\times$
0\farcs85/pixel (width $\times$ height).  The width 
of the slit was 2 pixels (0\farcs90) and its length was approximately 80\arcsec.
The final spectra were constructed from sets of 4 exposures, taken after
sequentially translating the detector array by 1/2 pixel with respect to
the spectrum produced by the fixed slit and grating. The final
reconstructed spectra have an effective resolving power of $\approx$
16,000 ($\Delta$V$_{\rm res. \ elt}$ $\approx$ 19~\kms). The slit was
oriented NS on July 5 and EW on July 10, with the Pistol Star near the
middle of the slit on both occasions. On August 3, two additional spectral
images were obtained, one with the slit at a position angle of 115$\arcdeg$ 
and centered at RA~=~17$^h$ 43$^m$ 5$\fs$16, Dec~=~-28$\arcdeg$ 48$\arcmin$ 45\farcs1 (B1950) to extend along
the ``barrel'' of the Pistol, the other with the slit at a position angle of 25$\arcdeg$ and
centered at RA~=~17$^h$ 43$^m$ 3$\fs$90, Dec~=~-28$\arcdeg$ 48$\arcmin$ 50\farcs3 (B1950) to lie along the ``handle'' of
the Pistol (see Fig.  2). The telescope was nodded well off of the
nebula to obtain ``blank sky'' images for subtraction. The on-source
integration times were 240 seconds on July 5, 768 seconds on July 10, and
240 and 480 seconds, respectively, for the August 3 positions. The seeing
was poor (1\farcs5 FWHM) on July 5 and was sub-arcsecond on the other
two nights. 

Data reduction was accomplished using the CGS4 on-line software
augmented off-line by the Figaro data reduction package maintained by
the Starlink Project. Bad pixel removal, flat-fielding, and coadding of
object-sky frame pairs were performed to produce the final spectral
images. The spectrum of the Pistol Star was extracted and then divided by
a similarly extracted spectrum of the nearby Quintuplet star, ``q3''\markcite{mgm94}
(nomenclature from\markcite{mon} Moneti et al.\ 1994), which is featureless in this
spectral region\markcite{f98} (see Figure 1 in F98).
 
The wavelength scales were determined by observing hydrogen and helium
emission lines from the planetary nebula NGC 6572 and converted to the
local standard of rest (LSR) frame\markcite{sch83} (Schneider et al.\ 1983). The final
reduced spectrum of the Pistol Star obtained on July 10 is shown in Figure
3. Portions of the wavelength-calibrated spectral images are shown in Figures 4a-d. 

\section{Analysis and Results}

The data have been analyzed in a variety of ways. First, we extracted photometry from the
emission-line and continuum images using DAOPHOT and the Pixon reconstructed data (which will
be discussed in detail in \S3.2), preferring the
latter method for extracting the estimate for the nebular flux. 
Second, we examined the continuum image (F190N) for possible companions
to the Pistol Star. Third, we estimated the 
extinction along the line of sight to the nebula and the mass of ionized gas in the nebula using the
continuum-subtracted image which was processed using the Pixon method.
Fourth, using this same image, we extracted point source photometry in order to 
obtain the 1.87~\micron\ flux excesses for the emission-line stars in the region. 
Lastly, we used the nebular spectra to extract the crucial velocity information which 
suggests an expanding nebula. We assume d$_{\rm GC}$~=~8,000~pc\markcite{reid93} (Reid 1993) 
throughout the paper.

\subsection{Photometry}

The photometry were performed using two methods. First, the Pixon deconvolution 
method (see below) was used to estimate the flux
from the Pistol Star in an infinite aperture from the originally-calibrated data. 
Second, the DAOPHOT package in IRAF\footnote
{IRAF is distributed by the National Optical Astronomy Observatories,
which are operated by the Association of Universities for Research
in Astronomy, Inc., under cooperative agreement with the National
Science Foundation.} was used to extract the number
of counts in a 0\farcs5 radius aperture from the newly-calibrated images; 
this value was then multiplied by 1.228 to simulate an infinite radius aperture. 
We find that this factor is appropriate, given the PSF profile measured in our data and shown in Figure 5. 
Results from both methods are shown in Table 3; they differ by
an insignificant amount. The flux at 1.90~\micron\
compares favorably with the value implied by fitting a curve
through the reddened spectral energy distribution in Figure 9 of 
F98, $\approx$ 3.9(10$^{-17}$) W cm$^{-2}$~\micron$^{-1}$; however,
this might just be a coincidence, given that the star is variable\markcite{glass99} (Glass et al.\ 1999).

\subsection{Pixon Image Reconstruction}

	The Pixon method's approach to deconvolution is to minimize
the chi-squared fit to the data of a minimum complexity, multiresolution
image model\markcite{pina93,puet95,puet99} (see Pina \& Puetter 1993; 
Puetter 1995; Puetter and Yahil
1999).  Relative to competing methods, use of a minimum complexity image
model results in a wide range of benefits, including increased spatial
resolution, decreased artifacts, and increased sensitivity to faint
sources otherwise hidden by these artifacts.  The Pixon method models 
the image as a local convolution of a
smoothing function and a pseudoimage.  The pseudoimage contains all
of the flux in the image while the smoothing functions decrease the
complexity of the image model by introducing at each point in the image
the maximum allowable correlation between image points.  A Pixon
reconstruction iterates between calculating the pseudoimage given a
set of smoothing kernels, and calculating the largest smoothing kernel
acceptable at each location in the image given the current pseudoimage.
Iteration proceeds until convergence, and no stopping criterion is required.
Usually only a few (2 or 3) iterations between pseudoimage
and smoothing kernel calculation are required for convergence.

The PSF used in the reconstruction was determined iteratively.  The data
profile around the Pistol Star itself was used as an initial guess for the PSF,
and this PSF was then used to perform an image reconstruction of the entire
field.  Next, using the relative brightnesses of additional stars in the field
and their determined centroids, a second, improved PSF was calculated from the
data by reversing the roles of image and PSF.  (Recall that a convolution
integral is invariant under the interchange of its components, i.e., the image
and the PSF.)  This produced a new PSF that was used to perform a second image
reconstruction of the field.  The second PSF was clearly a significant
improvement over the first PSF, but inspection of the second reconstruction
revealed the presence of weak field stars in the wings of the Pistol Star that
clearly contributed to artifacts in the wings of the PSF estimate. 
These were removed by hand by subtracting the current PSF
scaled to the brightness and position of the offending stars, and the
whole process was repeated. A new
reconstruction was performed, a third PSF determination was made, and a second
weak point source removal was performed by hand.  This last iteration produced
essentially no change in the PSF core and only minor changes in its wings, so
this final PSF was adopted. 

The Pixon image of the F187N (Pa-$\alpha$) mosaic is 
available in electronic form at {\it http://www.astro.ucla.edu/figer/papers.html}.
It has a dynamic range $\sim 3(10^5)$, well beyond the human perceptual capability, so the brighter
stars were saturated to allow the fainter stars plus nebula to take up most of
the displayed dynamic range.  The statistical significance of faint structures
in the image can be estimated from the residual errors.  Since a Pixon
reconstruction is made up of smooth Pixon elements, correlated on the scale of
the kernel function, the signal-to-noise ratio in the reconstruction is
measured per Pixon element and not per pixel.  Computed in this way, the
faintest parts of the nebula (well off its brighter parts) have ${\rm 
SNR}\ge 10$ with a typical ${\rm SNR}\sim 30$.

The most important feature of the reconstruction, however, is
the removal of the diffraction spikes of the stars.  The
diffraction sidelobes of the Pistol Star itself have been
successfully removed to about 1 part in $10^5$ in its immediate
vicinity, with the exception of a few artifacts discussed below.
Specifically, the reconstructed Pistol Star has nearly all of its
flux in a $3\times 3$ pixel box (0.106\arcsec\ radius) with the
central pixel containing 74\% of the total flux. This is to be
compared with the data, where 47\% of the flux extends beyond a
radius of 0.225\arcsec, 20\% outside of a radius of 0.375\arcsec,
and 9\% outside of a radius of 2.4\arcsec. However, a small amount
of remaining diffracted light from the Pistol Star does remain in
the reconstruction where the diffraction spikes run into the nebular
emission at a radius of 6.9\arcsec, too far for an accurate estimate
of the PSF.

One of the principle advantages of the Pixon method for crowded
fields is that it allows accurate photometry by disentangling the
emission for individual sources which has been blended together by
PSF blurring.  This is especially important in the case of the
Pistol Star where a significant fraction of the flux has blended
with the nebular emission and with that of nearby stars. Photometry
on the reconstructed image need only include the immediate area
around any object of interest (e.g., a few pixels on either side).
The relative photometric accuracy is limited in bright stars by systematic errors
to a precision of roughly 1\% of the total flux in individual point
sources.  For fainter stars, photon counting statistics and read
noise will dominate photometric errors. The 1\% error for bright
stars has contributions from two primary sources.  First, low level
artifacts arising from an imperfect PSF estimate can be seen
surrounding several of the brighter stars.  Second, inspection of
the residual errors (data minus the PSF convolved with the
reconstructed image) shows the presence of low level features from
an imperfectly removed diffraction pattern (first few Airy rings).
Nonetheless, the Pixon reconstruction recovers significantly more
flux than found by other methods, e.g., DAOPHOT. A number of artifacts are also obvious in the Pixon
reconstruction, i.e., ``holes'' dug out around some of the
stars. Their origin lies in the fact that not all stars are centered
in the data pixel grid in the same manner. Since the PSF used in the
reconstruction must necessarily assume some location for the point
source center, stars that are centered differently cannot be precisely
reconstructed.  The PSF centering selected here was for optimal
reconstruction of the Pistol Star since it had the largest
diffraction spikes.  Consequently, stars centered substantially
differently than the Pistol Star cannot be reduced to a point source.
To correctly fit the required centroid of the star, their central
peaks are broken into a number pixels.  However this makes the wings
of the star too broad.  If this profile were to be subtracted from
the data, a negative hole would result centered on the star. However
the Pixon method requires non-negative data.  So a somewhat broader
hole with flux values near zero intensity is created.  Such artifacts
can be removed only by performing a reconstruction in which the image
resolution is considerably higher than the data pixel grid.  Such a
reconstruction was not performed here since there was no reliable
method of obtaining the PSF on a commensurately fine scale.  Also
apparent in the Pixon reconstruction are low signal to noise
``blobby'' structures.  These occur especially around bright stars.
These features result from incompletely reconstructed starlight from
the bright stars due to errors in the determination of the
appropriate PSF for the stars in question.  These errors can arise
from a number of sources.  A primary error is due to different
centering of the PSF and the star in question.  Both of these error
sources are discussed above.  The presence of these ``blobby'' is
especially obvious around the Pistol Star itself, and near some of
the stars near the northern end of the western border of the frame.

Closer in (i.e. at small radii) to the stars showing surrounding
``blobby'' structure, the PSF is more completely removed in the
reconstruction.  There is a noticeable ``hole'' in emission, for
example, close to the Pistol Star (within a 1\farcs5 radius).
This emission has been reconcentrated back into the central pixel
of the Pistol Star in the reconstruction.  Of course weak nebular
emission might also be mistakenly recollected back into the Pistol
Star from this region since it is very difficult to distinguish
leaked stellar power (due to PSF misalignment) and low level nebular
emission so close to the Pistol Star.  Any such emission, however,
would affect the estimated flux for the Pistol Star at a very
low level ($\ll$1\%).

\subsection{Stellar image}

The HST image can be used to find coordinates for the Pistol Star. 
Our astrometry gives RA 17$^{\rm h}$ 43$^{\rm m}$ 4$\fs$85$\pm$0$\fs$1,  
DEC $-$28$^{\arcdeg}$ 48$^{\arcmin}$ 57$\farcs$23$\pm$1\arcsec\ (B1950), as compared to the coordinates
from Lang et al.\markcite{lang99} (1999) of RA 17$^{\rm h}$ 43$^{\rm m}$ 4$\fs$77$\pm$0$\fs$01,  
DEC $-$28$^{\arcdeg}$ 48$^{\arcmin}$ 57$\farcs$0$\pm$0$\farcs$1 (B1950). The errors for the HST 
position are dominated by systematic errors in determining the coordinates for the guide stars used
in setting the header values in the images. 
These astrometry agree with those in Nagata et al.\markcite{nag93} (1993): 
RA 17$^{\rm h}$ 43$^{\rm m}$ 4$\fs$8,  
DEC $-$28$^{\arcdeg}$ 48$^{\arcmin}$ 57$^{\arcsec}$ (B1950).

The high resolution of the images shows that the Pistol Star is single on scales greater
than $\approx$ 800 AU (0$\farcs$1), consistent with the much smaller limit of 110 AU
in F98.
Figure 6 shows the immediate vicinity of the Pistol Star in the F190N image. The first Airy ring 
is revealed, but no noticeable companions can be seen and the limit on the flux of any companion is
10\% of the flux of the Pistol Star. The contrast in brightness is greater than 10 to 1
between the brightest pixel on the peak and the brightest pixel in the Airy ring.
As further evidence for the star being
single, the Pistol Star provided the best point spread function in the field for the Pixon
deconvolution.

\subsection{Nebular morphology, flux, and mass}

Figures 1 and 2 show that the nebula completely surrounds the Pistol Star, although it
is quite faint on the southern and eastern portions. The nebula is dominated by roughly 
rectangular, concentric shells having sizes of 0.8$\times$1.2~pc and 0.4$\times$0.6~pc, and 
both major axes oriented
45$\arcdeg$ west of north. Note that their intensity characteristics differ
in that the brighter outer ring has strong brightness asymmetries.
There also appear to be two opposing lines of 
emission extending to the north and south of the star which are coincident with diffraction spikes from the
star. The northern line terminates in a bright
knot of emission (the ``northern knot''), where the nebula has its highest surface brightness flux density. 
Although this string of knots lies along a diffraction spike
associated with the Pistol Star, the flux in the string is $\approx$ 10 times greater than the flux
expected from the diffraction spike at its distance from the star, and the string does not show
the 3-line pattern of the spike. In addition, the string is clearly seen in the radiograph of Figure
13 in F98. 
There are also string-like features and other knots in the northern 
ridge (the ``barrel''; PA~=~115$\arcdeg$) and the western ridge (the ``handle''; PA~=~25$\arcdeg$).

The extinction along the line of sight to the nebula can be estimated by comparing the Pa-$\alpha$ flux to
free-free emission at 6 cm. After
subtracting flux from stars, we calculate F$_{\rm Pa-\alpha}$~=~
3.1(10$^{-19}$)$\pm$1.1(10$^{-20}$) W\,cm$^{-2}$ integrated over the nebula; the uncertainty is set
by the estimate of background flux. Lang, Goss, \& Wood\markcite{lang97} (1997) find 
0.5 Jy from the nebula at 6 cm. Using the relation between I$_{\rm Br-\gamma}$ and I$_{\rm 6 \, cm}$ 
in Wynn-Williams et al.\markcite{wynn} (1978), and
I$_{\rm Pa-\alpha}$/I$_{\rm Br-\gamma}$~=~12.5\markcite{hs87} (Hummer \& Storey 1987) for T$_e$~=~5,000 K and
n$_e$~=~500 cm$^{-3}$\markcite{lang97} (Lang et al.\ 1997), 
we find {\it A}$_{\rm 1.87 \, \mu m}$~=~3.8$\pm$0.3. This is equivalent to \AK~=~2.9$\pm$0.3
assuming {\it A} $\propto$ $\lambda^{-1.6}$\markcite{rrp} (Rieke, Rieke, \& Paul 1989). This value is
consistent with \AK~=~3.2$\pm$0.5 estimated in F98 for the Pistol Star.

We may now estimate the ionized gas mass:
\begin{equation}
M_{\rm ionized}~=~\rho_{\rm ionized} V,
\end{equation}
where $\rho_{\rm ionized}$ is the density of ionized gas and V is the emitting volume. We model the emitting volume
as a spherical shell with inner radius, r$_{\rm i}$, and outer radius, r$_{\rm o}$, i.e., 
\begin{equation}
V~=~{4 \over 3} \pi (r_o^3 - r_i^3) f,
\end{equation}
where f is the fraction of the shell which is ionized and accounts for incomplete ionization of
ejecta in the shell and an asymmetric external ionizing field.
The density is m$_{\rm p}$$<$n$_{\rm e}^2$$>$$^{1/2}$, where m$_{\rm p}$ is the mass of a proton, 
and $<$n$_{\rm e}^2$$>$$^{1/2}$ is deduced from the volume
emissivity for free-free emission, j$_{\rm \nu,ff}$, and the measured flux, F$_{\rm \nu}$:
\begin{equation}
F_{\nu}~=~{ j_{\nu,ff} V   \over 4 \pi d^2},
\end{equation}
where d is the distance between the Earth and the object. The volume emissivity in the Rayleigh-Jeans
limit is:
\begin{equation}
j_{\nu,ff}~=~0.018 {<n_e^2> \over T_e^{3/2}} {1 \over \nu^2} {g_{\nu,ff}} B_{\nu,T_e} \ \ ergs \, cm^{-3} \, s^{-1},
\end{equation}
where T$_{\rm e}$ is the electron temperature, g$_{\rm \nu,ff}$ is the gaunt 
factor, B$_{\rm \nu,T_{\rm e}}$ is the Planck function, and all quantities are in cgs units. 
Combining equations 3 and 4, we find:
\begin{equation}
<n_e^2>~=~{4 \pi d^2 \over V} {\nu^2 T_e^{3/2} \over 0.018 g_{\nu,ff} B_{\nu,T_e}} F_{\nu} \ \ cm^{-6}.
\end{equation}
Using this formulation, we find V~=~0.19~pc$^3$, $<$n$_{\rm e}^2$$>$$^{1/2}$~=~1,200 cm$^{-3}$,
and M$_{\rm ionized}$~=~5.6~\Msun, assuming r$_{\rm o}$~=~0.60~pc, r$_{\rm i}$~=~0.50~pc, 
T$_{\rm e}$~=~3,600 K, and g$_{\rm \nu,ff}$~=~4.2.  
The total gas in the ejection, M$_{\rm gas}$, is then M$_{\rm ionized}$/f
=~11~\Msun, assuming f~=~0.5. The value for the ionized fraction, f, 
is motivated by the fact that the ionized gas appears to be confined to the northern
side of the expanding ejecta\markcite{moneti} (Moneti et al.\ 1999). 
The adopted value of r$_{\rm o}$ is measured directly off the image, and r$_{\rm i}$ is
estimated from the apparent shell thickness.
We are also assuming that the thickness of the ionized layer is
equal to, or less than, the thickness of the shell.

The dust mass in the nebula can be estimated as follows:
\begin{equation}
M_{\rm dust}~=~
\left( F_{\nu}d^2 \over B_{\nu,T_{d}} \right)
\times A,
\end{equation}
where F$_{\nu}$ is the measured flux, T$_{\rm d}$ is the dust temperature, 
and {\it A} is the mass per effective cross-section of
the dust and is defined in Marston \& Dickens\markcite{mars88} (1988).

From Marston \& Dickens\markcite{mars88} (1988), 
{\it A} $\approx$ 0.013 g\,cm$^{-2}$ at 60~\micron, using their grain model. 
Taking F$_{60 \, \mu m}$~=~600 Jy and  T~=~173 K\markcite{simp97} (Simpson et al.\ 1997), 
we find, M$_{\rm dust}$~=~0.004~\Msun and M$_{\rm gas}$/M$_{\rm dust}$ $\approx$ 2,800
For any reasonable geometry, this dust mass implies A$_{\rm V}$ $\ll$ 1
for the whole nebula (consistent with section 4.3.3 in F98). 

\subsection{High resolution spectrum and the expanding ring}

Table 4 lists important parameters of the \ion{H}{1} 4-5 and \ion{He}{1} 4-5 lines in the
spectrum (Figure 3), along with values for those lines from the earlier spectrum in F98; notice
that the strength of the Br-$\alpha$ line is variable, confirming the behavior of the Br-$\gamma$
line (\ion{H}{1} 4-7, 2.166~\micron) reported in F98. 
The Br-$\alpha$ line again shows more emission on the blueward side, so that the centroid of emission is
shifted blueward from the peak by several~\kms. We interpret the asymmetry in the line as due to 
the opacity of the approaching (near) side of the wind which blocks emission from the
far side of the wind.

We now turn to the velocity field of the nebula, which we have mapped
for the first time using the long-slit Br-$\alpha$ spectra.
Figure 2 shows the slit positions observed, and Figures 4a-d show
the long-slit spectra; the velocity scale 
is normalized so that the zero-point corresponds to the radial velocity
of the Pistol Star, $\approx+130$~\kms\ (LSR).  The velocity field in
both the EW and NS slit positions traces an ellipse, consistent with 
an optically thin expanding nebula and similar to what is observed in planetary nebulae.  The emission 
is much stronger from the northern
and western part of the nebula, consistent with the emission seen in Figures 1 and 2.   
Referring back to Figure 2, the bright horizontal ridge of
emission is called the ``barrel'' and its emission (Figure 4c)
arises mostly from gas near the radial velocity of the
Pistol Star.  The patchy nature of the
emission is evident in Figure 4d, the long-slit spectrum of the ``handle'' of the Pistol Nebula.
In general, the redshifted emission is also much stronger than the blushifted
part of the ring, which can barely be seen in the images (also see Figure 6 in F98). The north-south 
long-slit spectrum (Figure 4a) appears very patchy, with the knot of bright
emission toward the North dominating the spectrum (see \S3.4). 

Most noteworthy is that our data clearly shows the
expansion signature and that the expansion ring is asymmetric.
The expansion velocity of the nebula cannot be extracted from the spectral images by simply
fitting an ellipse because the redshifted emission has a much flatter velocity pattern than
the blueshifted emission.  
The maximum velocity of the blueshifted gas is $-$60~\kms, while the maximum velocity of the
redshifted gas is $+$10~\kms, where both values are with respect to velocity of the Pistol Star.
We address the peculiar velocity field of the Pistol Nebula, and its possible origin, in \S4.

Our infrared spectra are consistent with the lower resolution radio data of Lang et al.\markcite{lang97} (1997). 
They find an average velocity in the H92$\alpha$ line of
115~\kms, roughly in the middle of the range of gas velocities found in this paper. They
were able to split the line profile into two components in the eastern (their Figure 7a) and the
southern (their Figure 11) portions of the nebula. In both cases, the velocities of the
fit lines are consistent with the two components we find in our data, i.e., V$_{\rm LSR}$
$\sim$ $+$140~\kms\ and V$_{\rm LSR}$ $\sim$ $+$70~\kms. In other regions, they fit a single
line having $\Delta$V$_{\rm FWHM}$ $\sim$ $+$60~\kms. Undoubtedly, these single line fits
are made to what we now see as two individual lines separated by about 70~\kms and having 
very small widths, $\sim$ 10~\kms. 

\subsection{Emission-line stars near the Pistol Nebula}

The continuum-subtracted Pa-$\alpha$ image reveals several emission-line stars (Figure 2), including stars
previously noted\markcite{fmm95,f99} (FMM95; FMM99).
Table 5 gives the locations and emission-line excesses for the hot stars in the field, where the photometry
were extracted from the Pixon reconstructed images. 

Equivalent widths can be estimated using equation 7. 
\begin{equation}
W_{1.87~\micron}~=~\frac{(F_{\rm F187N} - F_{\rm F190N})}{F_{\rm F190N}} \times \Delta \lambda,
\end{equation}
where the fluxes are in W\,cm$^{-2}$\,\micron$^{-1}$ and $\Delta\lambda$ is the 
FWHM of the F187N filter (0.0192~\micron); this equation assumes that 
the emission line falls completely within the filter bandwidth. Table 5 includes the
equivalent width estimates for the emission-line stars based upon this equation. As a check on this relation,
note that the equivalent width for the Pistol Star is within 5\% of the value given in 
F98, which was based on a direct spectral measurement. Of course, the
star is variable by about half a magnitude in the K-band\markcite{glass99} 
(Glass et al.\ 1999), so this coincidence might be fortuitous.

\subsubsection{Star \#235N and \#235S}

The current observations separate star \#235 into two emission-line stars having 
large values of F$_{\rm F187N}$/F$_{\rm F190N}$. FMM99\markcite{f99} observed the two
sources as a single star with a spectral type earlier than WC8 (``$<$WC8''). 
In theory, such
stars lack hydrogen, so any emission near 1.87~\micron\ should be attributable to 
helium lines. Is the 1.87~\micron\ flux from the two stars all due to \ion{He}{2}, 
as would be expected from $<$WC8 stars\markcite{fmn97} (Figer, McLean, \& Najarro 1997)?
We can estimate the expected flux from \ion{H}{1}, \ion{He}{1}, and 
\ion{He}{2} lines which fall in the F187N filter by scaling the equivalent widths
for similar in the K-band spectrum in FMM99\markcite{f99}. 

We do not expect any contribution from \ion{He}{1} because there are no such
lines in the K-band spectrum. The \ion{H}{1} contribution can be estimated by scaling
the maximum possible contribution to the 2.166~\micron\ line by Br-$\gamma$. 
The composite spectrum has W$_{2.166\, \mu m}$ $<$ 7 \AA. The two stars have similar continuum
levels, within 15\% at 1.9~\micron\ according to Table 5. Assuming that they have the same 
continuum flux ratios at 2.166~\micron, then the maximum equivalent width
for either component in the Br-$\gamma$ line  is $<$ 15 \AA. 
Hummer \& Storey\markcite{hs87} (1987) list recombination coefficients which indicate
W$_{\rm Pa-\alpha}$/W$_{\rm Br-\gamma}$~=~10.5 for T=10,000 K and n$_{\rm e}$=10$^4$ cm$^{-3}$\markcite{hs87}.
Thus, the flux excess at 1.87 um cannot be solely 
due to \ion{H}{1} 3-4, even if we attribute all the flux at 2.166~\micron\ to \ion{H}{1} 4-7.
As an aside, we argue that the 2.166~\micron\ feature (W $\approx$ 5 \AA) is
attributable to \ion{He}{2} 8-14, noting that the ratio of flux in that line to
flux in the 2.189~\micron\ line (\ion{He}{2} 7-10, 2.1891~\micron) is 1:4, identical to the ratio of 
recombination coefficients for the two associated transitions. 
Finally, it is likely that \ion{He}{2} lines contribute most of the flux in the F187N bandpass. 
W$_{\rm HeII\, 7-10}$~=~20 \AA\ in the composite K-band spectrum. Consulting Hummer \&
Storey\markcite{hs87} (1987) again, we find: 
W$_{\rm HeII\, 5-6, 1.8641\, \mu m}$/W$_{\rm HeII\, 7-10}$~=~9.6, 
W$_{\rm HeII\, 6-8, 1.8749\, \mu m}$/W$_{\rm HeII\, 7-10}$~=~2.5, and
W$_{\rm HeII\, 7-11, 1.8762\, \mu m}$/W$_{\rm HeII\, 7-10}$~=~0.6. 
The respective contributions to the flux in the F187N filter are: 
W$_{\rm HeII\, 5-6}$=190 \AA, 
W$_{\rm HeII\, 6-8}$=50 \AA, and 
W$_{\rm HeII\, 7-11}$=12 \AA.
All totalled, we estimate, W$_{\rm HeII\, 1.87\, \mu m}$~=~250 \AA, not too different
from the measured values in Table 5. Note that the three \ion{He}{2} lines
do not all fall in the same part of the F187N filter; however,
their line widths are broad enough to extend over most of the filter band
pass in all three cases\markcite{f99} (see FMM99).

We conclude that both stars are likely $<$WC8, given the coincidence of the following
facts: the above analysis indicates that the flux excess at 1.87~\micron\ is mostly due to 
\ion{He}{2} transitions, the composite K-band spectrum in\markcite{f99} FMM99 indicates very little hydrogen
in either star, and the excess emission at 3.09~\micron\ in the integrated light from 
both stars is ascribable to \ion{He}{2} 8-11\markcite{f99} (FMM99).

\subsubsection{Ionizing Flux}

Given the estimated ionizing flux from the various hot stars in the region, we can determine which
stars are contributing to the ionization of the nebula. Lang et al.\markcite{lang97} (1997)
estimate that the nebula is ionized by 5.6(10$^{48}$) \ion{H}{1} ionizing photons s$^{-1}$. 
Assuming that the Pistol Star emits $<$4.4(10$^{49}$) \ion{H}{1} ionizing photons s$^{-1}$ 
(see ``H'' model in F98) we find that it contributes $<$ 8\% of the ionizing photons at the most intense location
in the Pistol Nebula, assuming that the true distance between the hot stars and the nebula is 
the square root of two times the projected separation. An additional 85\% of the ionizing 
flux can be produced by star \#235N \& \#235S (44\%), \#241 (18\%), \#151 (13\%), and \#240 (9\%),
where a first guess of the relative contributions are in parentheses. The remainder of the flux 
could come from other hot stars in the region\markcite{f99} (Figer et al.\ 1999). 
This distribution of ionizing flux is consistent with the fact that the nebular emission is
most intense where the gas is closest to stars \#235N, \#235S, and \#241.
Note that if we use the ``L'' model in F98, then we would predict
that the Pistol Star contributes an insignificant amount of ionizing flux compared to other nearby
hot stars by the analysis above. In addition, the Pistol Star would produce seven orders of magnitude fewer
ionizing photons than what is required to ionize the nebula on its own.

\section{Discussion}

In this section, we argue that the Pistol Nebula is the most massive LBV ejecta known,
and that the expanding gas in the nebula has been significantly slowed by the
winds of nearby stars. We compare the nebula to other LBV nebulae and discuss
the possibility that the ambient magnetic field might impact the shape of the nebula.  

\subsection{The Pistol Nebula as LBV ejecta}

We interpret the Pistol Nebula as expanding LBV ejecta in the presence of 
massive stars with dense winds and a strong ambient magnetic field. 
The classic identifying characteristics of Luminous Blue Variable nebulae (LBVN) include chemical enhancement
relative to the local medium and a clear indication that gas is expanding from an
LBV\markcite{nota95} (Nota et al.\ 1995). 
Evidence for helium enhancement of the Pistol Nebula was given in Lang et al.\markcite{lang97} (1997)
and F98. If real, the enhancement is only slightly greater than
solar, but this is consistent with the suggestion that the Pistol Star is in the early
stages of an LBV phase\markcite{f98} (F98). 

The dynamical evidence presented in this paper, coupled with the
symmetry of the gas distribution with respect to the Pistol Star,
both on the sky (Figure 1) and in the position-velocity diagrams
(Figures 4a-d), shows that gas in the nebula is expanding from 
the Pistol Star.  We therefore suggest the following interpretation
of the observed velocities.  Gas in the Pistol Nebula was ejected
with a velocity similar to the current terminal velocity of the
stellar wind near the star, v$_{\rm wind}$~=~95~\kms\ (F98).  
The current expansion velocity of the nebula, v$_{\rm expansion}$ =
60~\kms, therefore indicates that the gas has been decelerated
due to interactions with the surrounding medium, a common occurrence
for LBVN\markcite{nota} (Nota et al.\ 1995).  The far side of the expanding shell has
been decelerated more than the near side, suggesting an interaction
between the expanding shell and the winds from other cluster stars
(see below).  The above values of v$_{\rm wind}$ and v$_{\rm expansion}$ 
constrain the expansion timescale for the outer nebula to lie between 
6,000 and 9,000 years.  The corresponding timescale for the inner shell
is half that of the outer shell.

The mass of gas in the nebula is much higher than any other LBVN, $\approx$ 10~\Msun. While this
might suggest that much of the material has been swept up from the ambient local medium, 
such material would not be enhanced in helium. In addition, the enhanced helium content (and large
nebular mass) cannot be swept up material from the winds of the WR stars in the
cluster\markcite{f98} (see \S6.2 in F98). Instead of seeing the large nebular mass as
a difficulty for the ejection hypothesis, it might be viewed as a natural consequence of the
particularly large mass of the progenitor. 
Indeed, we might expect that more massive stars will produce more massive LBVN\markcite{hd94}
(Humphreys \& Davidson 1994). 

Could such a large mass be decelerated by the winds of the nearby WC stars? Equation 8 is satisfied
if this is the case.

\begin{equation}
\rho_{\rm WC} {\rm v_{\rm WC}^2}~=~\rho_{\rm nebula} \delta \frac{\Delta V_{\rm expansion}}{\Delta t},
\end{equation}
where $\rho_{\rm WC}$ is the average density of the winds from the WC stars at the location
of the nebula, $\rho_{\rm nebula}$ is the density of gas in the expanding nebula, v$_{\rm WC}$
is the velocity of the winds, $\delta$ is the thickness of the expanding shell, $\Delta V_{\rm
expansion}$ is the velocity change of the shell, and $\Delta t$ is the time during which the
velocity changed. We assume an average distance between the WC stars and the nebula of 0.6~pc,
a total mass-loss rate in the WC star winds of $\approx$ 10$^{-4}$~\Msunyr, and a wind velocity
of 2,000~\kms. The change in velocity for the redshifted gas is 95~\kms\ $-$ 10~\kms~=~85~\kms. Using values in
\S3.4, we find $\Delta t$ $\approx$ 6,000 yr, roughly equivalent to our estimate of the 
expansion time. 

There is further evidence that the Pistol Nebula is circumstellar ejecta from the
Pistol star which is ionized by nearby hot stars. 
Moneti et al.\markcite{moneti} (1999) present results of a spectroscopic study of the Pistol and of
the cocoon stars in the Quintuplet Cluster.  From ISOCAM CVF 5---$17\micron$
spectroscopy, they find a nearly
spherical shell of hot dust centered about the Pistol Star, clearly indicating that the shell 
is stellar ejecta. The continuum images show that the dust shell extends far to the south of the
ionized portion of the nebula, consistent with the idea that the ionizing source is located outside
the shell to the north. The ISO emission-line images
confirm that most of the ionized material is along the northern border of the
shell, and the morphology in these images is very similar to that of the \HH\ region.  
The SWS spectrum of the nebula indicates a harder ionizing radiation than could be
provided by the Pistol Star, but which is consistent with ionization from
Wolf-Rayet stars in the Quintuplet Cluster.  
Taken together, all these data are consistent with the idea that the Pistol Nebula is 
circumstellar ejecta mostly ionized by very hot stars in the core of the Quintuplet Cluster.  

By most measures, the Pistol Nebula is typical, qualifying as a shell-type ejecta, like AG Car, 
according to Nota et al.\markcite{nota} (1995). Even the northern knot, and the trail of emission 
between it and the star, are reminiscent of features in the ``homunculus'' of $\eta$ Car\markcite{curr} (Currie 
et al.\ 1996).
Table 6 lists characteristics of LBVN, as adapted from Nota et al.\markcite{nota} (1995), with a
new entry for the Pistol Nebula. Within errors, most of the values for the 
listed parameters are within in the range of values for other LBVN, 
except for gas temperature and gas mass.
The gas temperature is consistent with the relatively low temperatures for
\ion{H}{2} regions in the Galactic Center\markcite{lang97} (c.f. Lang et al.\ 1997). 
So, the total nebular mass is the only truly extraordinary characteristic of the Pistol Nebula,
being over a factor of two greater than the mass of the next most massive LBVN. 

\subsection{Shaping of the Pistol Nebula by the ambient magnetic field?}

The quasi-rectangular morphology of the nebula is not consistent 
with perfectly spherical expansion, although that is not unusual for 
the ejecta from LBVs\markcite{nota} (Nota et al.\ 1995), presumably because, in many 
cases, the ejections are themselves irregular and anisotropic.  
In the case of the Pistol Nebula, the ambient magnetic field 
in which the nebula is apparently embedded may have an important 
influence on the nebular geometry.  Three considerations lead us to 
this conclusion.

	First, the nebula is bounded by two nonthermal radio 
filaments (NTF's) which delineate the magnetic field direction (see Figure 7).  These 
filaments appear to be associated with the bundle of NTF's 
constituting the Radio Arc\markcite{yzm} (Yusef-Zadeh \& Morris 1987), and the 
uniformity of the filament orientation indicates that the magnetic 
field is perpendicular to the Galactic plane throughout this region\markcite{morr94,chan99} 
(indeed, Morris 1994 [see also Chandran et al.\ 1999] has argued that 
the magnetic field is perpendicular to the Galactic plane throughout 
the central 100 - 150~pc).  

The second reason to suspect the importance of the magnetic field for 
shaping the geometry of the nebula is the elongation and 
orientation of the nebula: it is elongated along the magnetic field, as 
would be expected if the ambient field were strong enough to 
constrain the expansion in the direction perpendicular to the field.  
On the other hand, the Pistol Nebula has a roughly rectangular 
morphology, with the long sides parallel to the field, as defined by 
the bounding NTF's, while the shape to be expected for expansion of 
a spherical piston into a strong magnetic field has been predicted to be 
more oval, at least in the earlier stages of the expansion\markcite{inse91,ferr91} (Insertis and 
Rees 1991; Ferriere, Mac Low \& Zweibel 1991).  In any case, the 
minor axis dimension and the axial ratio of the nebula should be 
consistent with the ambient magnetic field strength, if indeed the 
field is responsible for shaping the nebula.  Noting the apparent 
rigidity of the nonthermal filaments, especially where they interact 
with molecular clouds or \ion{H}{2} regions, Yusef-Zadeh and Morris\markcite{yzm,morr94} (1987, 
see also Morris 1994) have argued that the field strength within the 
filaments, and perhaps throughout the Galactic center region, is at 
least a milligauss.  

To illustrate that such a strong field can have an 
important dynamical effect on the ejecta from the Pistol Star, we 
equate the ram pressure of the expanding wind, $\rho V_{\rm w}^2$, with 
the magnetic pressure, B$^2$/8$\pi$, in order to estimate the stagnation radius, 
r$_{\rm m}$, in the direction perpendicular to the 
magnetic field.  With $\rho$~=~$\Mdot/4\pi r_{\rm m}^2 V_{\rm w}$, 
where~\Mdot\ is the mean mass loss rate during an ejection event, we obtain:
\begin{equation}
r_{\rm m} ~=~
0.12 ~pc ~ 
\left( { \Mdot      \over 10^{-4}~\Msun~yr^{-1} } \right)^{1/2} ~
\left( { V_{\rm expansion}  \over 100~\kms              } \right)^{1/2} ~
\left( { 1~mG       \over B                     } \right).
\end{equation}

The semi-minor axis of the nebula is about 10\arcsec, or 0.39~pc at the distance to the Galactic center.  This 
distance can be made to agree with the calculated stagnation radius 
if: 1) the initial ejection velocity was substantially higher than the 
current expansion velocity, 2) the mass ejection rate during 
an outburst substantially exceeds 10$^{-4}$~\Msunyr, and/or the B-field is weaker than 1 mG.

Either of these possibilities are plausible, given the values in Table 6; however, it is difficult to see
how the strength of the magnetic field can be significantly below 1 mG\markcite{yzm} 
(Yusef-Zadeh \& Morris 1987). It seems likely that the mass loss rate during an eruption is substantially
enhanced over the time-averaged value.  Assuming, for example, that  
r$_{\rm outer}$~=~0.60~pc, r$_{\rm inner}$~=~0.50~pc, v$_{\rm eruption}$
=~95~\kms, and M$_{\rm eruption}$~=~11~\Msun, 
we find \Mdot~=~0.02~\Msunyr, giving r$_{\rm m}$~=~1.7~pc.  This
value of r$_{\rm m}$, and that derived from an assumed constant mass loss rate,
bracket the observed value; given the uncertainties in the various 
parameters, we regard the observed minor axis size of the Pistol Nebula 
to be consistent with the possibility that it has been magnetically shaped.

The third consideration is the brightness discontinuity of the bounding 
filaments across the boundary of the nebula.  
The brightness of each of the bounding filaments changes 
along their lengths from one side of the nebula to the other by a factor of $\approx$2, indicating that their 
emissivity has undergone a discontinuity somewhere along the 
boundary. In addition, the northern filament becomes two filaments on the eastern side of the nebula.
This not only supports the notion that the bounding NTF's 
are interacting with the nebula, but it may point to the nebular 
boundary as a source or a sink for relativistic electrons.  In their 
model for the principal NTF's of the Radio Arc, Serabyn and\markcite{morr94} Morris 
(1994) hypothesized that the \ion{H}{2} region G0.18$-$0.04 is the source of 
relativistic electrons for those filaments because of the pronounced 
brightness discontinuities they suffer where they encounter that \ion{H}{2} region.  
The acceleration mechanism in this model is magnetic field line 
reconnection at the interface between the \ion{H}{2} region and the 
ambient interstellar medium.  A related model can be applied to 
G0.15-0.05, except that in this case, there is no molecular cloud 
underlying the \ion{H}{2} region.  The model of Rosner \& Bodo\markcite{rosn96} (1996) 
seems particularly apt because it is based upon a stellar wind from a 
hot star.  These authors suggest that the electron acceleration takes 
place at the wind's termination shock.  However, they consider winds 
with velocities $\sim$1,000~\kms, or more than an order of 
magnitude larger than that of the Pistol Star, so their results will have to 
be scaled accordingly.  

\section{Conclusions}

Our new imaging and spectroscopic data suggest that the Pistol Star 
has recently ($\lesssim$ 10$^4$ yrs) ejected a massive 
circumstellar nebula, distributed in two shells 
centered on the Pistol Star, which is being photoionized by nearby hot stars in the Quintuplet Cluster. 
The ejected mass is $\approx$ 10~\Msun, over a factor of two times the most massive
ejecta from an LBV.
The nebula has been ejected in the presence of a strong magnetic field and powerful winds
of nearby hot stars, offering an opportunity to test wind-wind interaction models and
models of magnetically confined outflows. 

\acknowledgements

We gratefully acknowledge the late Chris Skinner of STScI for 
providing assistance in performing the HST observations.
We thank Christine Ritchie of STScI for calibrating the data. We thank Zolt Levay of
STScI for producing the color composite image in Figure 1. 
We thank L. E. Bergeron of STScI for ``cleaning'' the images of bad pixels using his ``imclean'' IDL
script. We thank Antonella Nota for providing the original Latex code for Table 6. We thank Tony Marston for
discussions concerning the impact of massive stars on their local medium. We thank Andrea Moneti
for useful discussions concerning his ISO results on the Pistol Star.
We also thank Rodger Thompson, Marcia Rieke, and the NICMOS team for providing assistance in reducing our data. 
Support for this work was provided by NASA
through grant number GO-07364.01-96A from the Space Telescope Science Institute, which is operated by
AURA, Inc., under NASA contract NAS5-26555. The CGS4 spectra were obtained as 
part of the UKIRT Service Programme. UKIRT
is operated by the Joint Astronomy Centre on behalf of the United Kingdom
Particle Physics and Astronomy Research Council. 

\clearpage
\small
\begin{deluxetable}{crr}
\tablewidth{0pt}
\tablecaption{Summary of Observations}
\tablehead{
\colhead{Position} &
\colhead{F187N Dataset} & \colhead{F190N Dataset}
}
\startdata
N & N44T01K6	 & N44T01K9		 \nl
E & N44T01KC	 & N44T01KF			   \nl
S & N44T01KJ 	& N44T01KM		           \nl
W & N44T01KP 	& N44T01KS		           \nl
\enddata
\end{deluxetable}

\clearpage
\small
\begin{deluxetable}{ccc}
\tablewidth{0pt}
\tablecaption{Calibration Images}
\tablehead{
\colhead{Calibration} &
\colhead{Original Calibration} & \colhead{Final Calibration}
}
\startdata
Mask & H4214599N & H4214599N		           \nl
Noise & H4216218N & H4216218N \nl
Linearity & H3V1404NN & H4N1654BN \nl
Dark & H7B1528ON & HAD12034N \nl
Flat\tablenotemark{a} & H1S1337HN/H1S1337JN & I3A1616AN/I3A1616BN 		           \nl
Photometric table & H7M1005JN & I3G1220AN \nl
Illum\tablenotemark{a} & H2413241N/H2413243N	 & H2413241N/H2413243N			   \nl
\tablenotetext{a}{F187N/F190N.}
\enddata
\end{deluxetable}

\clearpage
\small
\begin{deluxetable}{lcc}
\tablewidth{0pt}
\tablecaption{Photometry}
\tablehead{
\colhead{Method} & \colhead{F$_{\rm F187N}$} & \colhead{F$_{\rm F190N}$} \\
\colhead {}      & \colhead{W cm$^{-2}$~\micron$^{-1}$} & \colhead{W cm$^{-2}$~\micron$^{-1}$}
}
\startdata
Pixon\tablenotemark{a} & 6.22(10$^{-17}$)) & 4.12(10$^{-17}$)         \nl
DAOPHOT\tablenotemark{b} & 6.24(10$^{-17}$) & 4.14(10$^{-17}$)         \nl
\tablenotetext{a}{A 0\farcs375 radius aperture was used to sum the counts.
The zero-points are 4.336(10$^{-18}$) ergs~cm$^{-2}$~\AA$^{-1}$~DN$^{-1}$ for F187N and
4.293(10$^{-18}$) ergs~cm$^{-2}$~\AA$^{-1}$~DN$^{-1}$ for F190N.}
\tablenotetext{b}{A 0\farcs5 radius aperture was used to sum the 
counts, and the result was multiplied by 1.228, as indicated by our analysis of the PSF in the images.
The zero-points are 3.537(10$^{-18}$) ergs~cm$^{-2}$~\AA$^{-1}$~DN$^{-1}$ for F187N and
3.519(10$^{-18}$) ergs~cm$^{-2}$~\AA$^{-1}$~DN$^{-1}$ for F190N.}
\enddata
\end{deluxetable}

\clearpage
\begin{deluxetable}{lcrrrr}
\small
\tablewidth{0pt}
\tablecaption{Spectral Lines}
\tablehead{ 
& 
& 
\colhead{$\lambda_{\rm vac}$} & 
\colhead{V$_{\rm LSR}$} & 
\colhead{$\Delta$V$_{\rm FWHM}$} & 
\colhead{W$_{\lambda}$} \\
\colhead{Source\tablenotemark{a}} &  
\colhead{Species} &
\colhead{\AA} &
\colhead{\kms} & 
\colhead{\kms} & 
\colhead{\AA} 
}
\startdata
this work	& \ion{He}{1} 4-5 & 40490\phantom{.}\phn & 107$\pm$15 	& 110$\pm$30 & 3.0$\pm$0.5 	\nl
F98 		& \nodata 		& \nodata 			&  81$\pm$15	& \nodata 	 & 3.1$\pm$0.4\tablenotemark{b} \nl
this work 	& \ion{H}{1} 4-5 	& 40522.4 			& 126$\pm$5\phn 	& 125$\pm$10 & 199$\pm$10 \nl
F98 		& \nodata 		& \nodata 			&  117$\pm$15	& 118$\pm$15 & 150$\pm$5	\nl
\enddata
\tablenotetext{a}{F98 data were taken from Figer et al.\ (1998). Note that V$_{\rm LOS}$ given in that
paper is equal to V$_{\rm LSR}$ $-$ 10~\kms.}
\tablenotetext{b}{The value in F98 is in error.}
\tablecomments{Values were extracted from from UT 1998 July 10 data. Uncertainties are given in parentheses.}
\end{deluxetable}

\clearpage
\tiny
\begin{deluxetable}{ccccrrrrrrr}
\tablewidth{0pt}
\tablecaption{Emission-line Stars}
\tablehead{
\colhead{ID} & \colhead{Other\tablenotemark{a}} & \colhead{Sp. Type\tablenotemark{a}} & \colhead{R.A.\tablenotemark{b}}
& \colhead{DEC.\tablenotemark{b}} & \colhead{I$_{\rm F187N}$/I$_{\rm F190N}$\tablenotemark{c}}  
& \colhead{F$_{\rm F190N}$} & \colhead{W$_{1.87\, \mu m}$\tablenotemark{d}} & \colhead{log(Q)} & \colhead{Log(L/\Lsun)} \\
\colhead {} & \colhead {}      & \colhead {} & \colhead{s} & \colhead{\arcsec} & \colhead {} 
& \colhead{W cm$^{-2}$~\micron$^{-1}$} & \colhead{\AA} & \colhead{} & \colhead{} }
\startdata
1 & \#235S\tablenotemark{e} & \nodata & 4.78 &	35.51  & 2.69 & 1.17(10$^{-18}$)$\pm$2.9(10$^{-20}$) & 332 & 49.8 & 6.14 \nl
2 & \#235N\tablenotemark{e} & \nodata & 4.77	& 33.34 & 2.27 & 1.04(10$^{-18}$)$\pm$2.8(10$^{-20}$) & 249 & 49.8 & 6.12 \nl
3 & \#76 & WC9 & 5.16	& 71.91 & 1.95 & 1.01(10$^{-18}$)$\pm$2.7(10$^{-20}$) & 188 & 49.5 & 5.81 \nl
4 & \nodata & \nodata & 4.39 & 	46.95 & 1.52 & 1.91(10$^{-19}$)$\pm$1.2(10$^{-20}$) & 104 & \nodata & \nodata \nl
5 & \#134 & LBV & 4.85 & 57.23 & 1.50 & 4.12(10$^{-17}$)$\pm$4.1(10$^{-19}$) & 99 & $<$41.5\tablenotemark{f} & 
6.61\tablenotemark{f} \nl
6 & \nodata & \nodata & 3.81 &	48.74 & 1.24 & 1.10(10$^{-19}$)$\pm$8.7(10$^{-21}$) & 48 & \nodata & \nodata \nl
7 & \#151 & WC8 & 4.43  &  54.31 & 1.21 & 1.41(10$^{-18}$)$\pm$3.2(10$^{-20}$) & 43 & 49.5 & 5.96 \nl
8 & \nodata & \nodata & 6.31 &	45.18 & 1.13 & 3.07(10$^{-18}$)$\pm$4.7(10$^{-20}$) & 28 & \nodata & \nodata \nl
\tablenotetext{a}{Designations and spectral types are from FMM99. \#134 is the 
Pistol Star.}
\tablenotetext{b}{Coordinates are plus R.A. 17$^{\rm h}$ 43$^{\rm m}$ and DEC. $-$28\arcdeg\ 58\arcmin\ (B1950).
The error is 0$\fs$01 and 0$\farcs$1.}
\tablenotetext{c}{I~=~F$\Delta$, see text. Photometry taken from Pixon reconstructed data.}
\tablenotetext{d}{Equivalent widths are estimated as discussed in the text.}
\tablenotetext{e}{\#235 was first detected as one object with a spectral type earlier than WC8 (FMM99).}
\tablenotetext{f}{These values are from the ``L'' model in F98. Note that the ``H'' model
would give log(L/\Lsun)~=~7.20, and log(Q)~=~49.6.}
\tablecomments{Stars were selected if F$_{\rm F187N}$/F$_{\rm F190N}$~$>$~1.10. Some targets were
rejected if their flux ratio appeared to be confused by nebulosity. Lyman continuum fluxes and luminosities 
for the WC stars are estimated using the new photometry in this paper and the method described 
in FMM99.}
\enddata
\end{deluxetable}

\clearpage
\tiny
\begin{deluxetable}{lcrrrrrrrr} 
\tablewidth{0pt}
\tablecaption{LBV Nebulae\tablenotemark{a}}
\tablehead{
\colhead{}       &
\colhead{}& 
\colhead{}    & 
\colhead{Nebular}  &
\colhead{}    & 
\colhead{Dynamical}  &
\colhead{}  & 
\colhead{} & 
\colhead{}  &
\colhead{} 
\\
\colhead{Star}       &
\colhead{log(L/\Lsun)}& 
\colhead{V$_{\rm wind}$}    & 
\colhead{size}  &
\colhead{v$_{\rm gas}$}    & 
\colhead{time}  &
\colhead{M$_{\rm gas}$}  & 
\colhead{M$_{\rm dust}$} & 
\colhead{N$_{\rm e}$}  &
\colhead{T$_{\rm e}$} 
\\
\colhead{}       &
\colhead{} & 
\colhead{(\kms)}    & 
\colhead{(pc)}  &
\colhead{(\kms)}    & 
\colhead{(10$^4$ yr)}  &
\colhead{(\Msun)}  & 
\colhead{(\Msun)} & 
\colhead{(cm$^{-3}$)}  &
\colhead{(K)} \\
\colhead{(1)}       &
\colhead{(2)} & 
\colhead{(3)}    & 
\colhead{(4)}  &
\colhead{(5)}    & 
\colhead{(6)}  &
\colhead{(7)}  & 
\colhead{(8)} & 
\colhead{(9)}  &
\colhead{(10)}
}
\startdata
Pistol Star& 6.6 &  95 & 0.8x1.2 & 60 & 0.6$-$0.9\phn\phn &  11\phn\phn &  0.003\phn &  1200  & 3600\tablenotemark{b} \\
AG Car     &   6.2   &  80$-$250 & 1.1x1.0  & 70      &0.84\phn   & 4.2\phn\phn\phn & 0.013\phn     & 800      &9000\\
R 127      &   6.0   &$\sim$150& 1.9x2.2  & 28      & 4.0\phn\phn    & 3.1\phn\phn\phn & \nodata   & 1000     &\nodata\\
$\eta$ Car &   6.5\tablenotemark{c} & \nodata & 0.2      & 600     &  0.015 &  \nodata   &  0.01\phn\phn& \nodata  &\nodata\\ 
He 3-519   &   4.8   &   365   & 2.28     & 61      &1.8\phn\phn    & 2.0\phn\phn\phn  & 0.0066    & 300      &8000\\
S 119      &\nodata  & \nodata& 1.9x2.1  & 25      & 5.0\phn\phn & 1.7\phn\phn\phn &  \nodata  & 800      &\nodata\\ 
WRA 751    &   5.8   & \nodata& 0.8      & 20$-$60 & \nodata  & \nodata       &  \nodata   & \nodata &\nodata\\
HD168625   &   5.4   &    250 & 0.06      & 20      & 0.3\phn\phn    & 0.03\phn\phn& 0.0003    & 630      &\nodata\\ 
HR Car     &   5.7   & 145$-$170 & 0.98     & 50      & 1.0\phn\phn  & 2.1\phn\phn\phn & 0.0010  & 600      &$<$12500   \\
R143       &  6.0    & \nodata &   3.5    & \nodata  & \nodata       &  \nodata & \nodata &  \nodata & \nodata\\ 
P Cyg      &   5.2   & 206     & 0.2      &140      & 0.01$-$0.01    & 0.0092    & \nodata  & 1000     &5000   \\ 
\enddata
\tablenotetext{a}{All values are from Table 1 in Nota et al.\markcite{nota95} (1995), and references
therein, unless noted otherwise.}
\tablenotetext{b}{From Lang et al.\markcite{lang97} (1997).}
\tablenotetext{c}{$\eta$ Car is probably multiple\markcite{dam97} (Damineli, Conti, \& Lopes 1997).}
\tablecomments{The columns indicate: star name (1), luminosity (2), velocity of the stellar wind (3), 
size of the nebula (4),
gas expansion velocity (5), age of the nebula (6), ionized gas mass (7), dust mass (8), electron density (9), and
electron temperature (10).}
\end{deluxetable}

\newpage

\figurenum{1}
\figcaption[pistrgb.ps]
{Color composite of the F187N minus F190N image (red), the F190N image (blue),
and the F187N image (green). The reddish hue reveals ionized hydrogen gas recombining through the
n=4 to n=3 transition. The plate scale for all images is 0\farcs076 pixel$^{-1}$ in x and 0\farcs075
pixel$^{-1}$ in y. Pixel (253,254.5) corresponds to the center of the mosaic; see text and 
Figure 2 caption for coordinates.}

\figurenum{2}
\figcaption[diff.ps]
{Negative greyscale image of the continuum-subtracted flux in the narrow band F187N filter
($\lambda$~=~1.87~\micron). The Pistol Nebula, Pistol Star (near center), and several emission-line
stars can clearly be seen. The longest emission ridge near the top of the
figure is the ``barrel''  while the short emission ridge to the right is the ``handle.''
Nomenclature from FMM99 is used for previously-identified emission-line
stars; the numbering from Table 4 is used for newly-identified emission-line stars. 
The pair of stars labelled, ``235,'' are referred to as 235N and 235S in the text.
Slit positions on the nebula are indicated and labelled according to their positional 
angle on the sky; ``+'' symbols are at the slit centers.}

\figurenum{3}
\figcaption[pspec.ps]
{High-resolution spectrum (R $\approx$ 16,000) of the Pistol Star near the Bracket-$\alpha$ line.
Zero velocity corresponds to the velocity of peak line emission in the star.}

\figurenum{4a}
\figcaption[specns.ps]
{Long-slit spectrum of the Pistol Star and associated
nebula with the slit oriented NS as indicated on Figure 2.  
North is up and positive velocity indicates a redshift.  
The velocity zero point corresponds to the peak of the Bracett-$\alpha$
line in the Pistol Star, $\approx$$+130$~\kms.  The origin of the spatial scale
(y-axis) for this figure and 4b-d is indicated by a ``+'' symbol on 
the slit positions shown on Figure 2.  Notice the elliptical envelope
of the emission, consistent with an expanding nebula.}

\figurenum{4b}
\figcaption[specew.ps]
{Long-slit spectrum of the Pistol Star and nebula
with slit oriented EW (Figure 2); east is up.  Axes as in Figure 4a.  Notice
the nearly complete ellipse of emission, consistent with an expanding nebula.}

\figurenum{4c}
\figcaption[spec115.ps]
{Long-slit spectrum of the ``barrel'' of the nebula (Figure 2).
The slit was oriented with PA~=~25\arcdeg. Southeast is up.}

\figurenum{4d}
\figcaption[spec25.ps]
{Long-slit spectrum of the ``barrel'' of the nebula (Figure 2).
The slit was oriented with PA~=~115\arcdeg. Northeast is up.  The
spectrum near the top of the image is due to an emission line star.}

\figurenum{5}
\figcaption[psf_power.ps]
{Plot of enclosed energy as a function of aperture radius for PSF in the F190N filter.}

\figurenum{6}
\figcaption[pairy.ps]
{Log stretch of central 1$\arcsec$ $\times$ 1$\arcsec$ region surrounding the Pistol Star. The first
Airy ring can clearly be seen, and there is no evidence of a bright companion. The brightest pixel
on the peak has 10 times the flux of the brightest pixel in the Airy ring.}

\figurenum{7}
\figcaption[new6.ps]
{Log greyscale plot of 6 cm radiograph. The origin is set at the location of the Pistol Star. The
nonthermal filaments can clearly be seen, bounding the Pistol Nebula on its northeast and southwest
sides.}

\newpage

\begin{figure}
\end{figure}

\begin{figure}
\hspace*{0.75in} 
\plotone{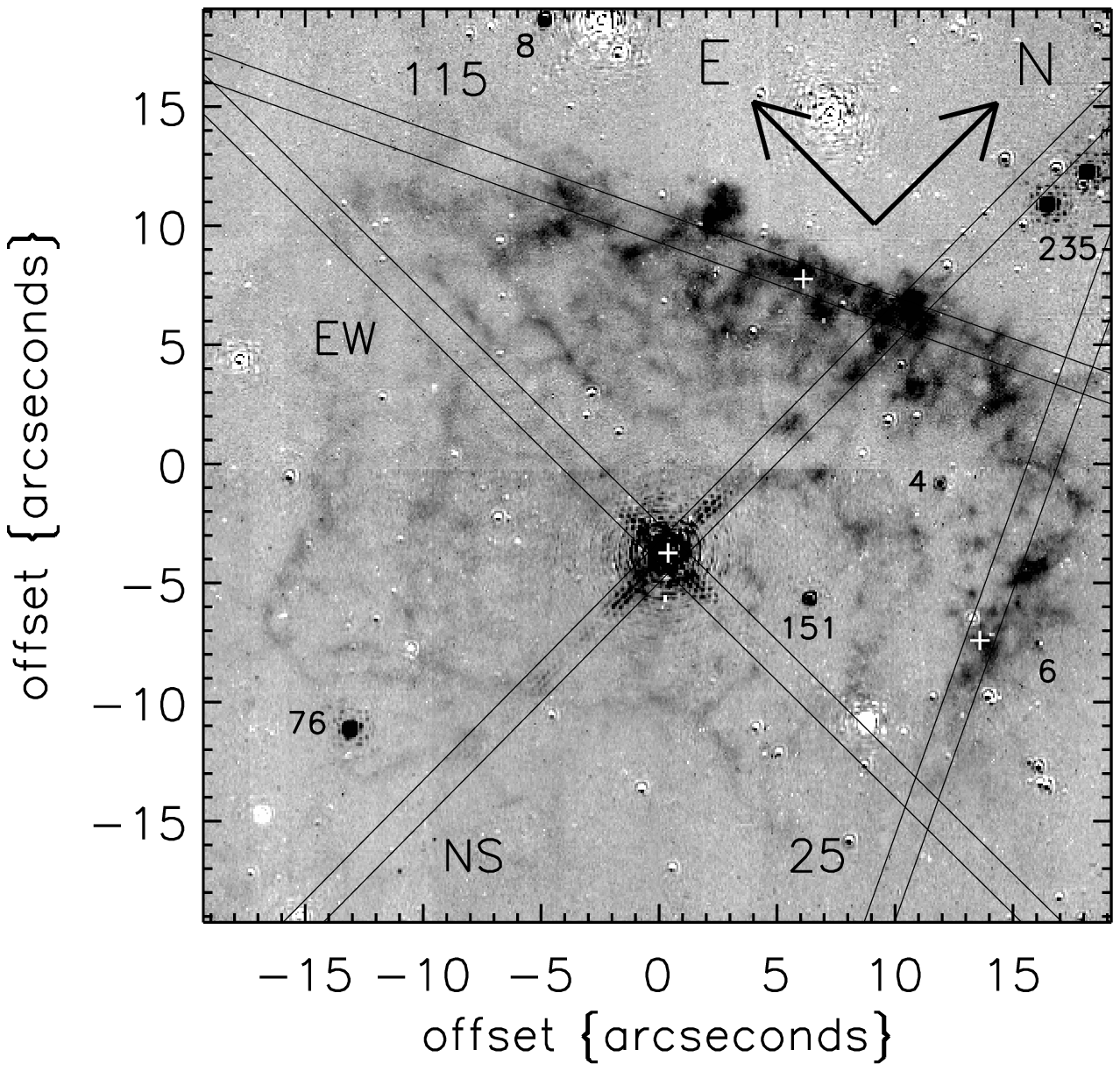}
\hspace*{4.5in} 
\vskip .2in
Figure 2
\end{figure}

\begin{figure}
\hspace*{0.75in} 
\plotone{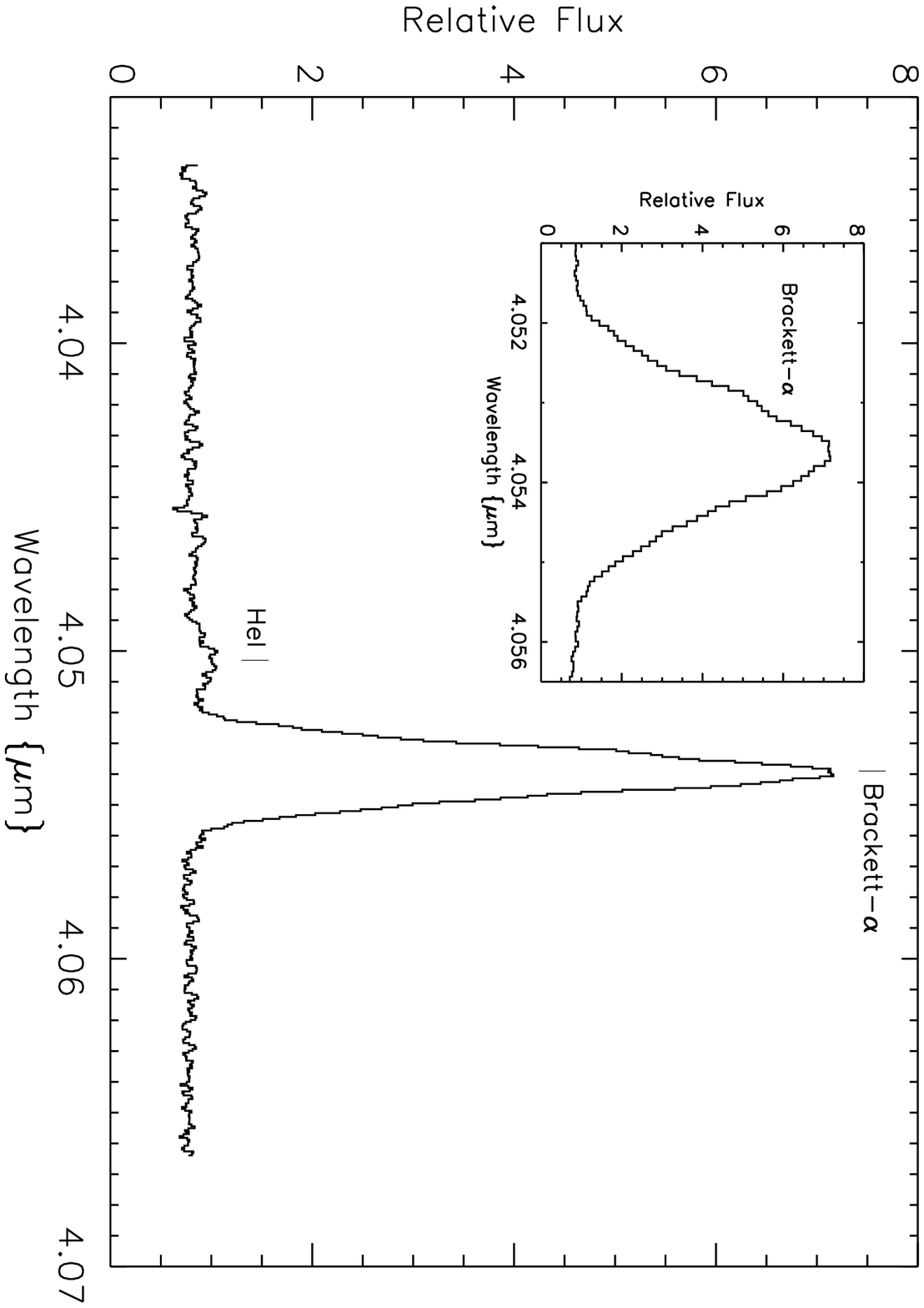}
\hspace*{4.5in} 
\vskip .2in
Figure 3
\end{figure}

\begin{figure}
\hspace*{1.in} 
\plotone{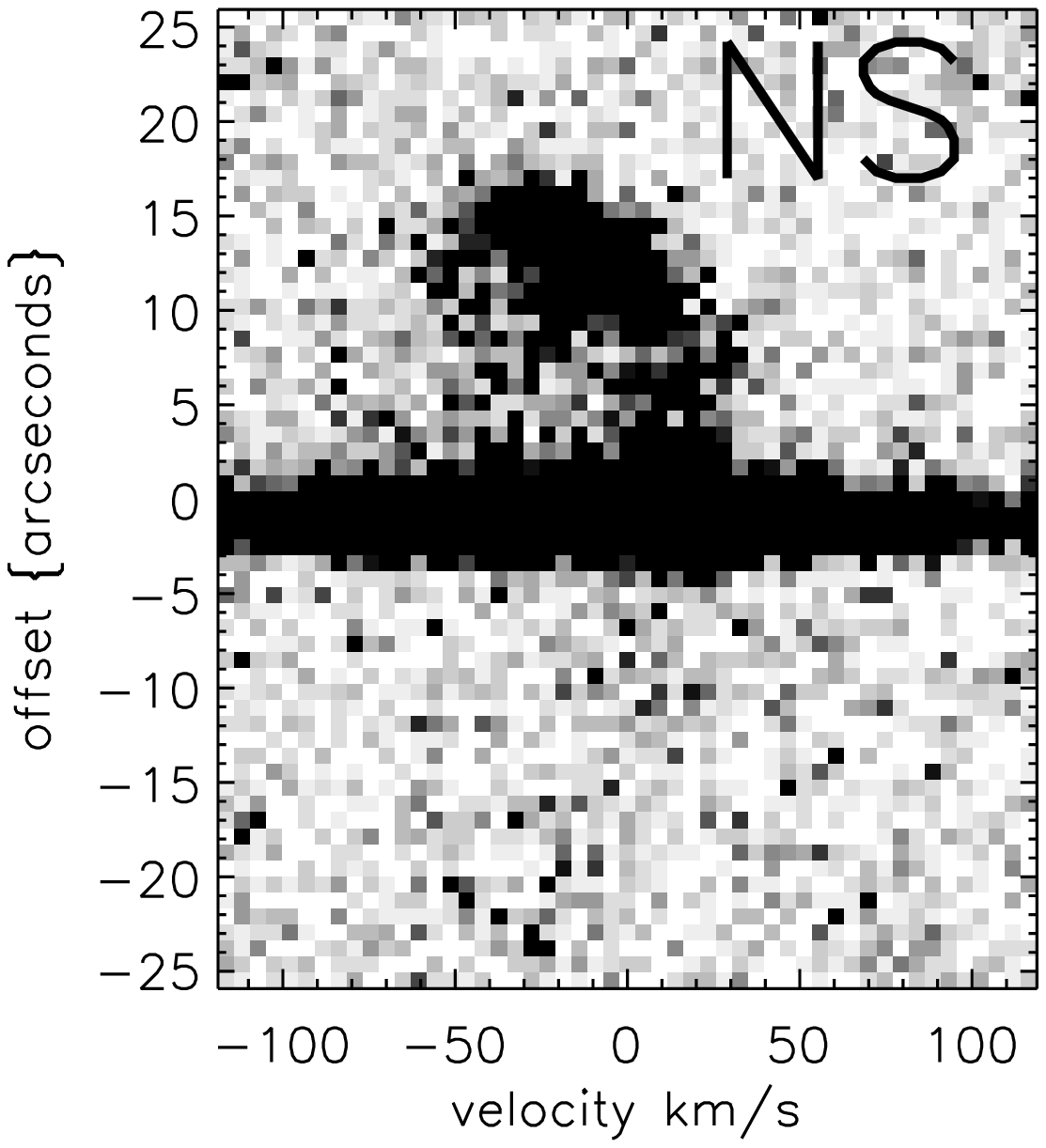}
\hspace*{4.5in} 
\vskip .2in
Figure 4a
\end{figure}

\begin{figure}
\hspace*{1.in} 
\plotone{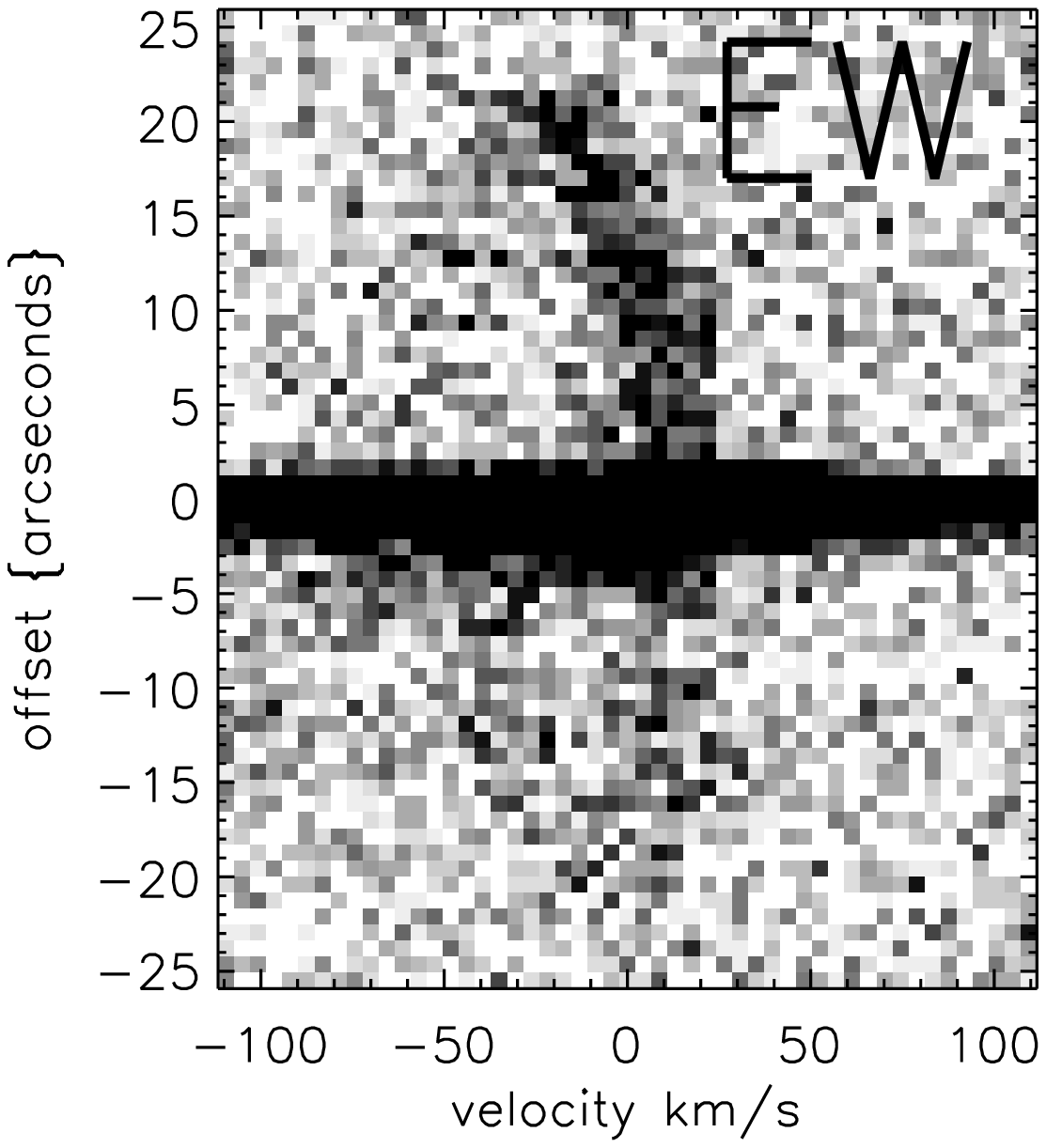}
\hspace*{4.5in} 
\vskip .2in
Figure 4b
\end{figure}

\begin{figure}
\hspace*{1.in} 
\plotone{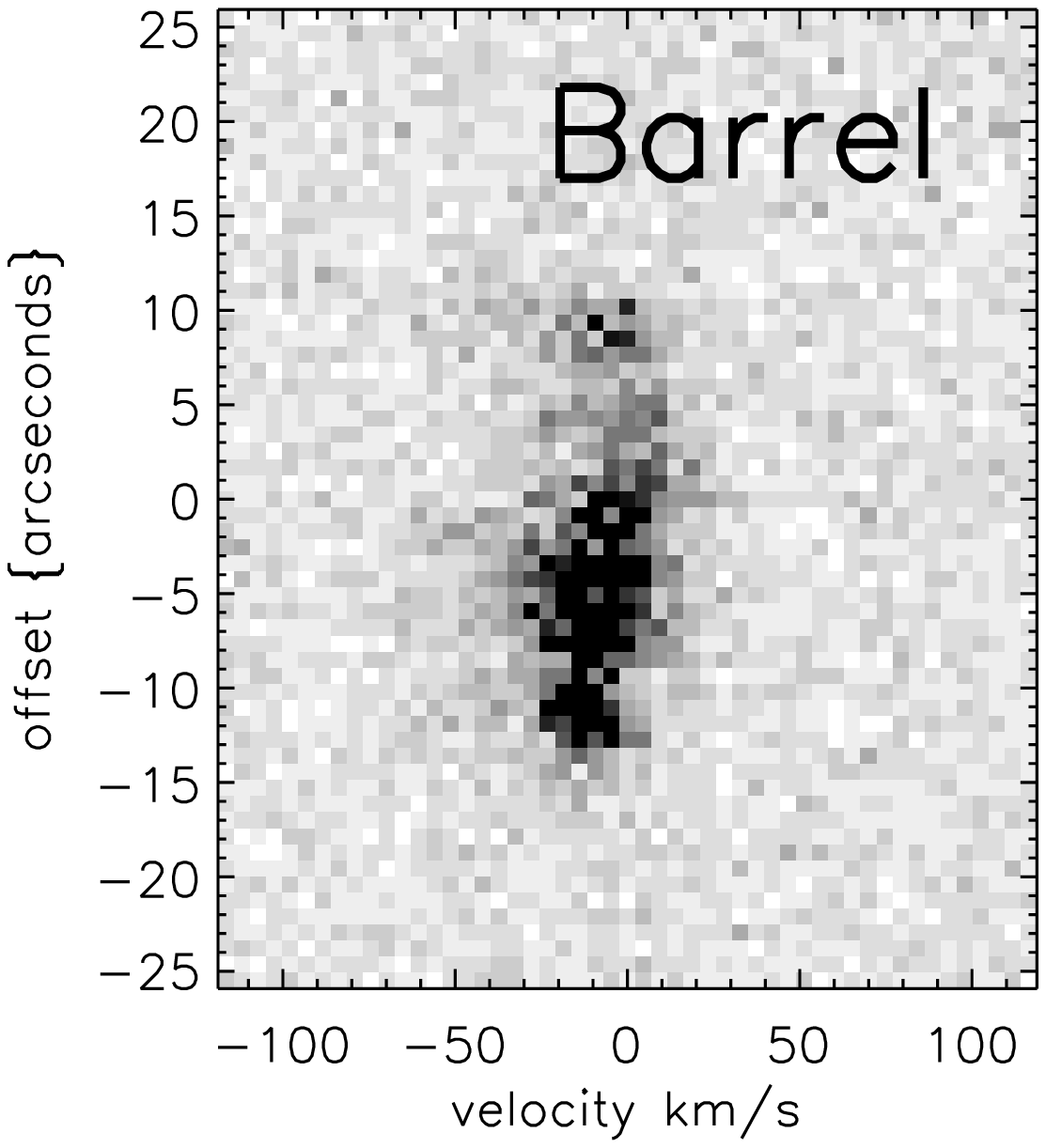}
\hspace*{4.5in} 
\vskip .2in
Figure 4c
\end{figure}

\begin{figure}
\hspace*{1.in} 
\plotone{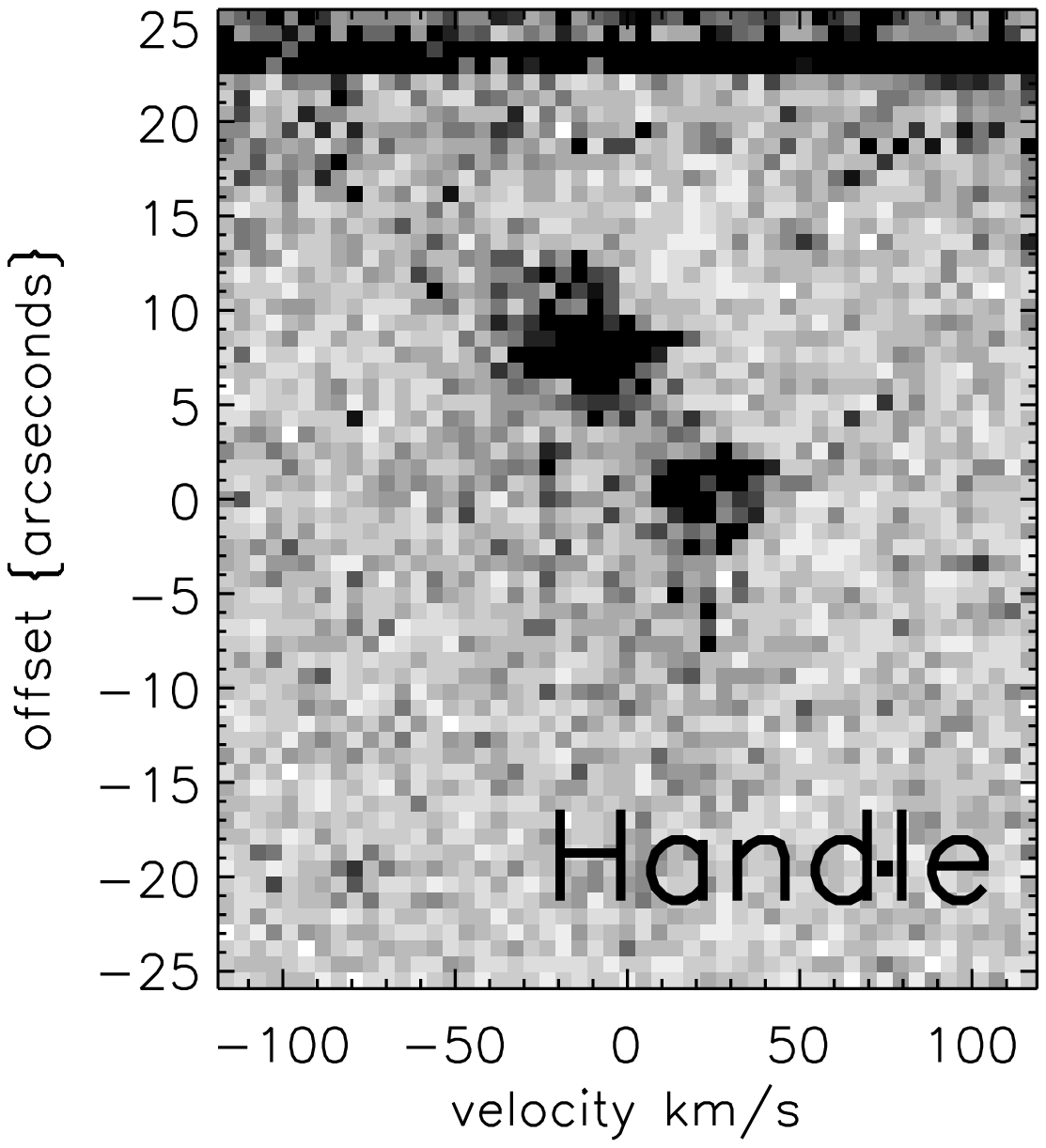}
\hspace*{4.5in} 
\vskip .2in
Figure 4d
\end{figure}

\begin{figure}
\hspace*{0.75in} 
\plotone{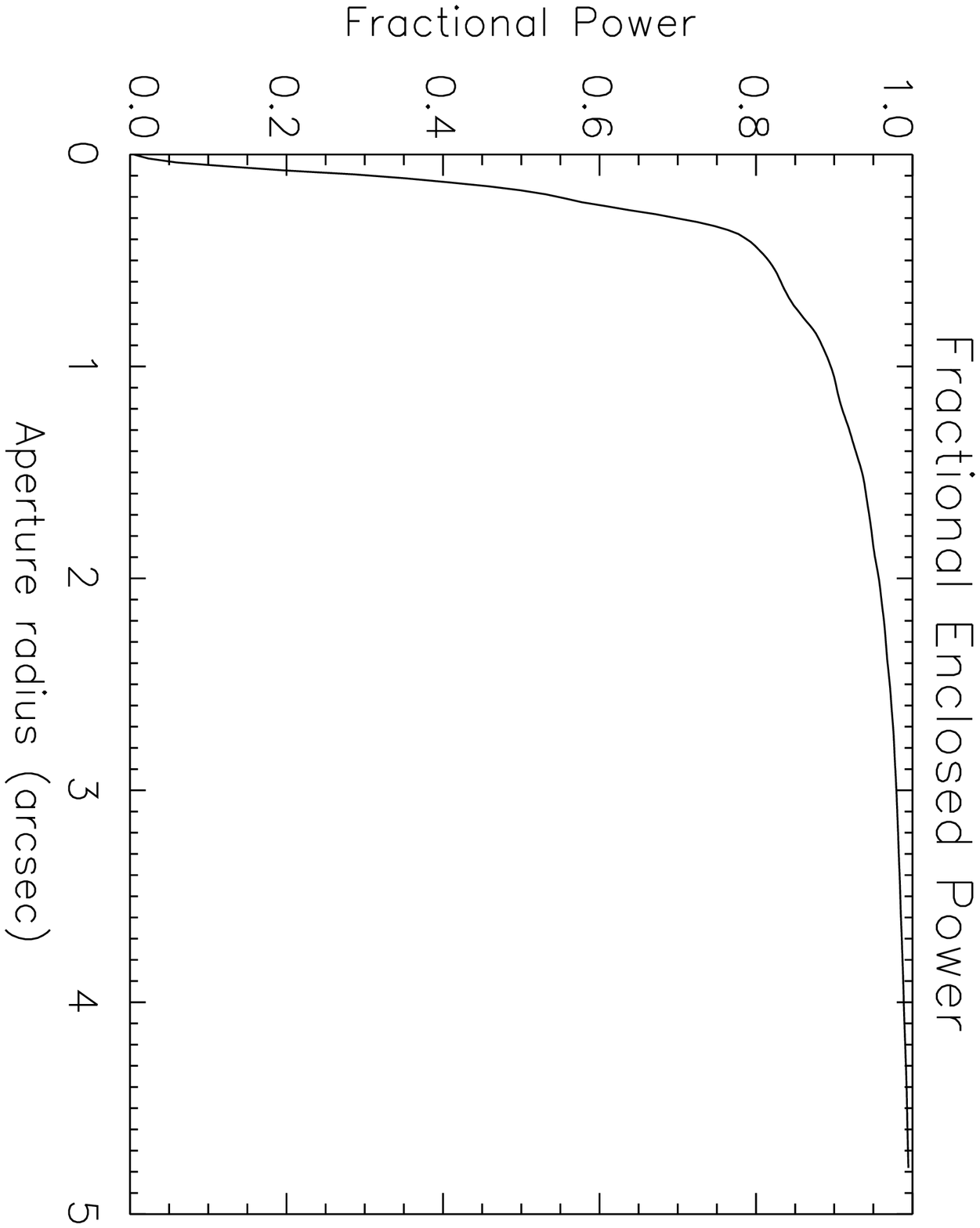}
\hspace*{4.5in} 
\vskip .2in
Figure 5
\end{figure}

\begin{figure}
\hspace*{0.75in} 
\plotone{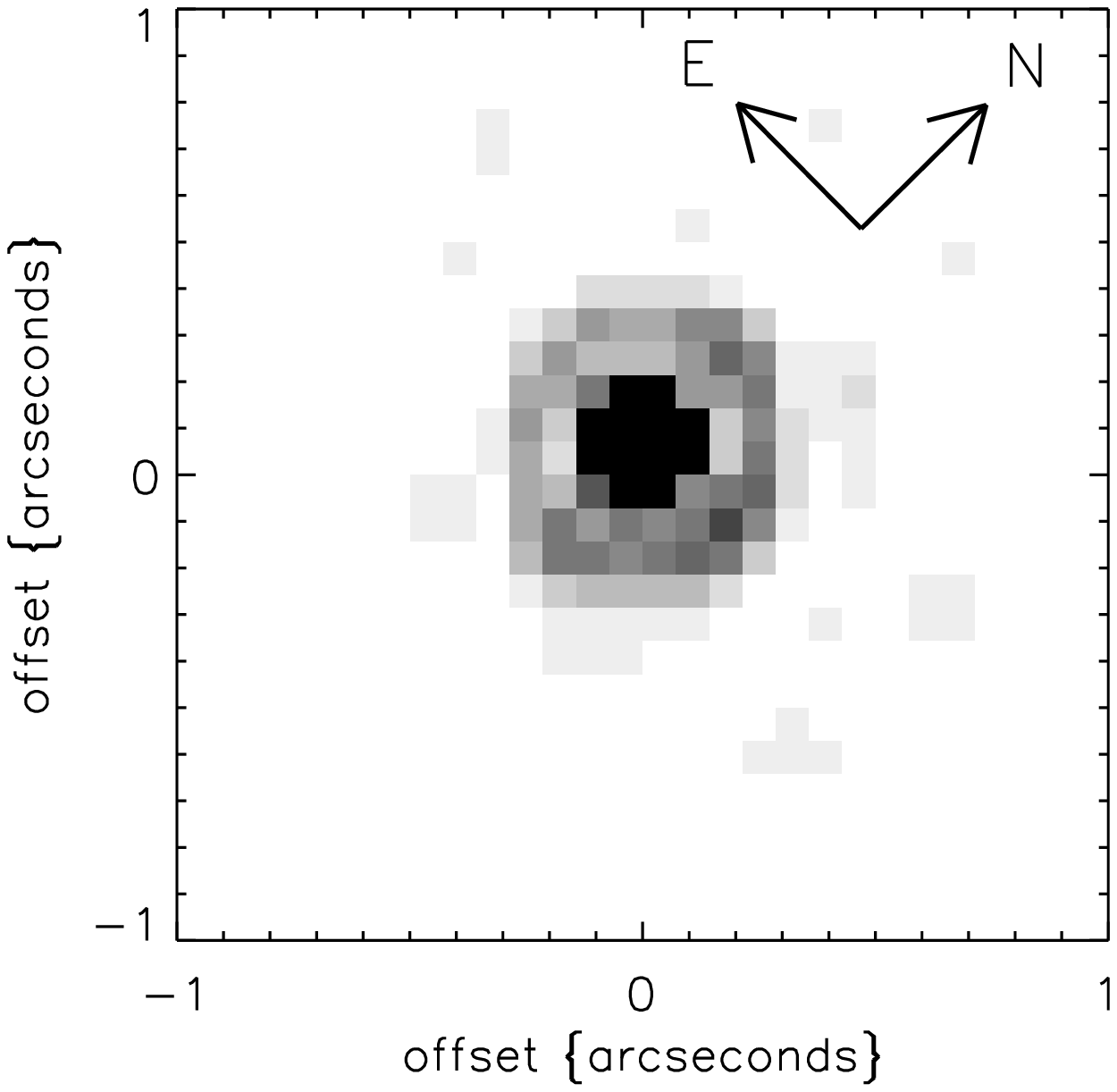}
\hspace*{4.5in} 
\vskip .2in
Figure 6
\end{figure}

\begin{figure}
\hspace*{0.75in} 
\plotone{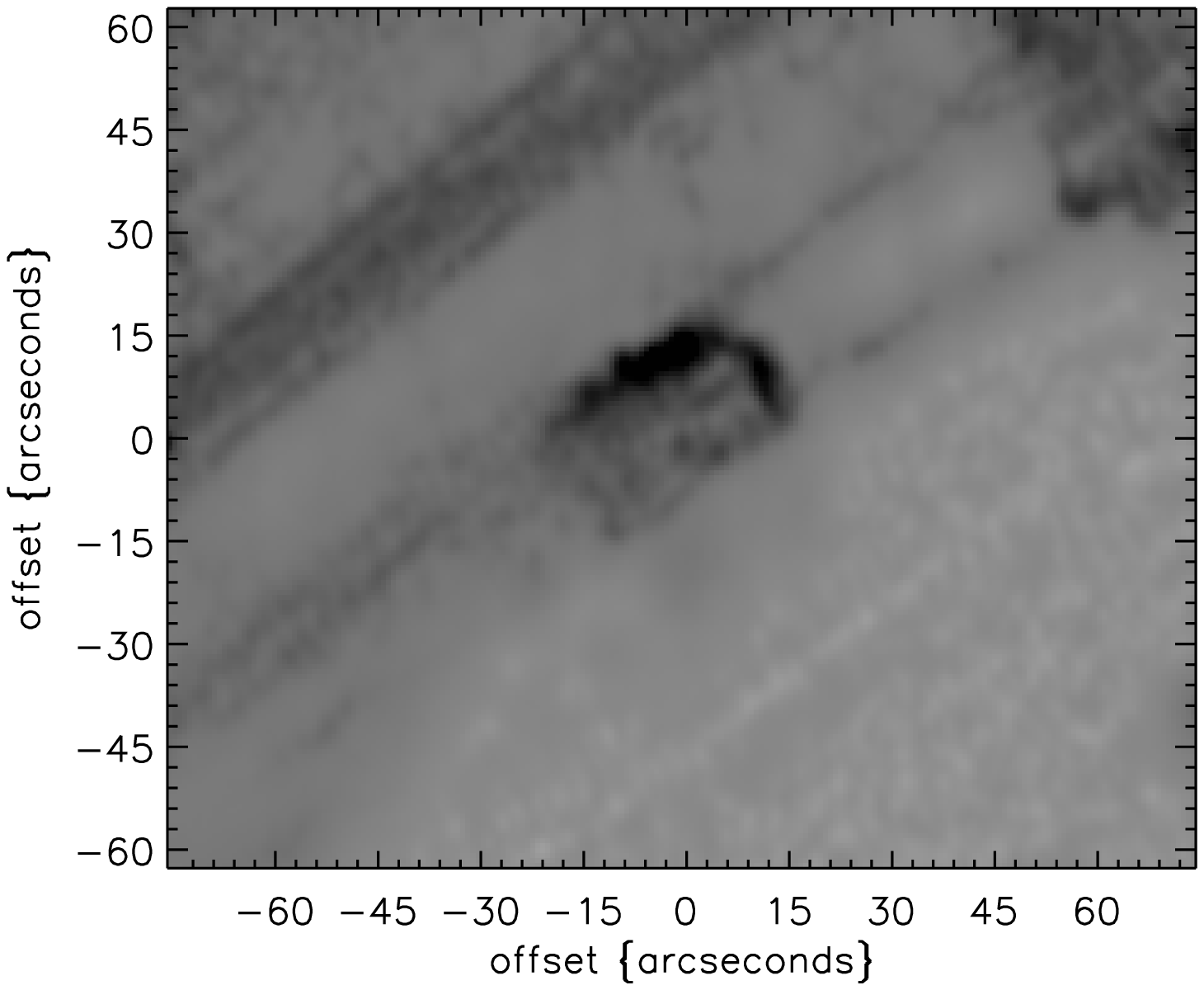}
\hspace*{4.5in} 
\vskip .2in
Figure 7
\end{figure}

\end{document}